\documentclass{article}
\usepackage{spconf,amsmath,amssymb,graphicx,booktabs}
\usepackage[T1]{fontenc}
\usepackage[hidelinks]{hyperref}
\usepackage{needspace}
\title{ReKV: Overlapping Quantized KV Transfer and Suffix Recomputation}
\name{Guangjian Dong}
\address{New York University}

\begin{document}
\ninept
\maketitle

\begin{abstract}
Prefill--decode disaggregation moves a prompt's key--value (KV) cache between workers before the first token can be generated. Reducing this handoff to either cache compression or recomputation misses a useful choice: which layers to transmit, at what precision, and which layers to execute again on the decode worker. ReKV transfers a quantized prefix KV cache and one BF16 boundary state, then reconstructs a contiguous suffix while the prefix is in flight. In a model-shaped synthetic benchmark on cross-node H100s, the Q8 configuration meeting separately evaluated quality criteria reduces 8K handoff latency by 2.36$\times$, 2.84$\times$, and 3.65$\times$ over BF16 transfer for Qwen2.5-7B, Qwen2.5-3B, and Qwen3-8B. It is also faster than the best measured transfer-only or recompute-only action at those points. A separate real-weight BF16 recomputation experiment reduces prefill-to-token-return time by 1.16--1.63$\times$. The preferred action changes with prompt length and link bandwidth.
\end{abstract}


\section{Introduction}

Prefill--decode (P/D) disaggregation assigns prompt processing and autoregressive generation to different workers. The two phases place different demands on the GPU~\cite{sarathi}, but the separation also makes the prompt KV cache a network payload~\cite{splitwise,distserve,mooncake,dejavu}. The decode worker needs the prompt state before it can continue autoregressive generation. For long prompts, moving that state can consume a substantial fraction of time to first token (TTFT).

The handoff admits two familiar ways to reduce transferred bytes. Low-bit representation sends less KV data at the cost of quantization work and possible changes in the output distribution~\cite{kivi,kvquant,cachegen}. Recomputation sends less state but spends decode-side compute to rebuild it. These choices interact: quantizing a shorter transmitted prefix changes both the link occupancy and the amount of work available to overlap with suffix execution. A policy selected for one model or link need not remain best at another prompt length.

ReKV studies this interaction for a fresh prompt handoff between separate prefill and decode GPUs. It selects a layer boundary and a precision for the prefix KV state. The prefill worker sends the hidden state at the boundary first. The decode worker uses that state to execute the remaining Transformer layers while the quantized prefix KV state is transferred and restored. This construction reproduces the suffix computation from the BF16 boundary state and quantizes only the transmitted prefix.

The evaluated decision couples prefix precision with a contiguous suffix boundary. Prior systems study mixed-precision KV transfer~\cite{spectrumkv} and partial KV restoration with overlapping I/O~\cite{kvpr,cacheflow}. We compare the joint action on the same cross-node P/D path against full BF16 transfer, low-bit transfer without recomputation, and BF16 suffix recomputation without quantization. At 8K tokens, ReKV with Q8 is faster than each single-mechanism alternative for three model configurations. A faster L40S link instead favors direct transfer for Qwen2.5-7B, while Qwen2.5-3B at 512 tokens favors Q8 transfer without recomputation.

\begin{figure*}[t]
  \centering
  \includegraphics[width=0.90\textwidth]{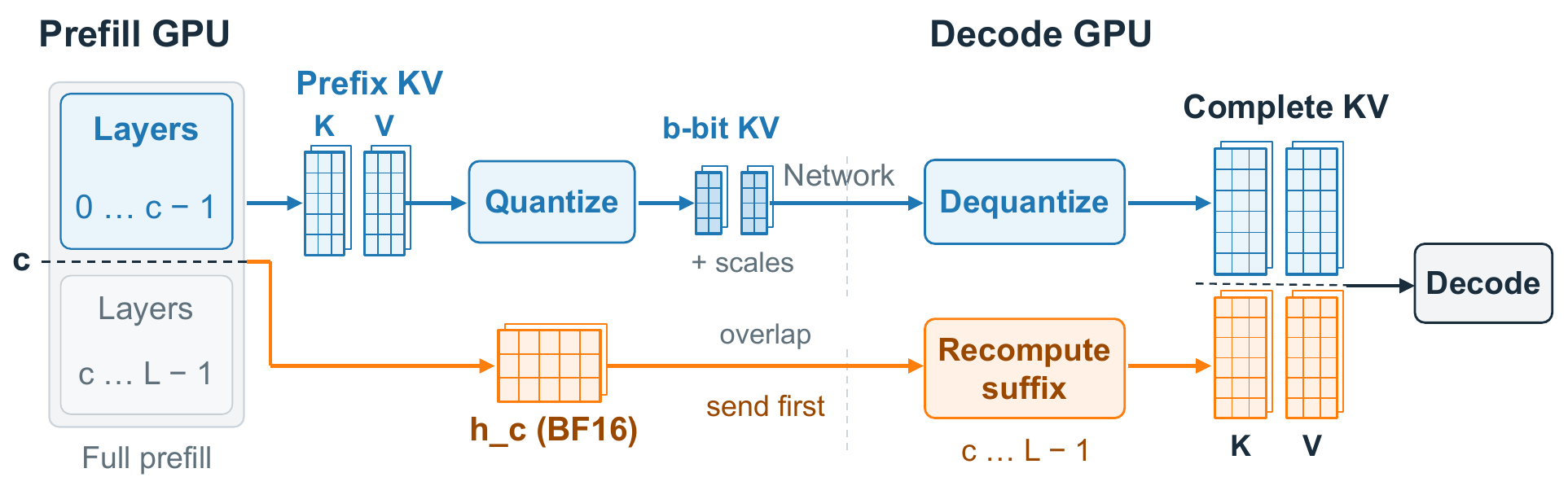}
  \caption{ReKV handoff at a cut $c$. The prefill worker sends the BF16 boundary state $h_c$ before the quantized KV state of layers $0,\ldots,c-1$. On the decode worker, suffix execution and prefix transfer proceed concurrently; the restored prefix and recomputed suffix form the complete cache needed for decoding. The blue and orange paths denote transferred and recomputed state, respectively.}
  \label{fig:mechanism}
\end{figure*}

\section{Joint Handoff Design}

\subsection{Why the boundary is necessary}

Consider a grouped-query-attention model~\cite{gqa} with hidden width $d$, $n_h$ query heads, $n_{kv}$ KV heads, head dimension $d_h$, and prompt length $S$. With $e$ bytes per element, a boundary hidden tensor and one layer's KV state require
\begin{align}
 B_H &= Sde, \\
 B_{KV} &= 2Sn_{kv}d_he.
 \label{eq:layerbytes}
\end{align}
The factor of two accounts for both keys and values. Using $d=n_hd_h$ gives
\begin{equation}
  \frac{B_H}{B_{KV}}=\frac{n_h}{2n_{kv}}.
  \label{eq:ratio}
\end{equation}
For the grouped-query-attention configurations studied here, the ratio is 3.5 in Qwen2.5-7B and 2.0 in Llama-3.1-8B~\cite{qwen25,llama3}.

\noindent\textbf{Proposition 1 (per-layer hidden-state dominance).}
For equal element precision and the same path bandwidth $W>0$, if $n_h\geq 2n_{kv}$, transferring a hidden state and projecting its K and V cannot be faster than transferring that layer's KV state in the bandwidth-plus-projection cost model.
\par\noindent\emph{Proof.} Let $T_{proj}\geq0$ be the time to produce K and V from the hidden tensor. The difference between the two costs is
\begin{equation}
 \begin{split}
 \Delta T &= \frac{B_H-B_{KV}}{W}+T_{proj} \\
 &= \frac{Sd_he}{W}(n_h-2n_{kv})+T_{proj}.
 \end{split}
\end{equation}
Both terms are nonnegative under the stated condition. The difference is strictly positive if $n_h>2n_{kv}$ or $T_{proj}>0$. \hfill$\square$

A single boundary state instead enables the receiver to execute a suffix of complete layers.

\subsection{Action and schedule}

For a model with $L$ layers, let $c$ be the first recomputed layer, so the suffix length is $r=L-c$. The prefill worker retains KV state for layers $0,\ldots,c-1$ and the BF16 residual state $h_c$ at the cut. It quantizes the prefix KV tensors in groups of 64 elements along the last dimension, sending packed payloads and FP16 scales. The decode worker receives $h_c$, starts layers $c,\ldots,L-1$ on a compute stream, and receives and dequantizes the prefix on a separate path. It waits for both branches before decoding (Fig.~\ref{fig:mechanism}). The boundary actions are transfer-only ($r=0$) and full recomputation ($r=L$); the latter sends token IDs rather than a boundary tensor.

For an interior cut, one boundary replaces $r$ layers of transmitted KV state. In BF16, the resulting byte count is
\begin{equation}
 B_{16}(c)=cB_{KV}+B_H,\qquad 0<c<L.
 \label{eq:bytes}
\end{equation}
Relative to full transfer, the saved bytes are $rB_{KV}-B_H$. Thus a BF16 cut reduces communication exactly when
\begin{equation}
 r>\frac{B_H}{B_{KV}}=\frac{n_h}{2n_{kv}}.
 \label{eq:amortize}
\end{equation}
For Qwen2.5-7B, at least four recomputed layers amortize the boundary tensor in bytes. This condition concerns communication volume; suffix execution still contributes to latency.

Let $E=2Sn_{kv}d_h$ be the number of KV elements per layer. With $b\in\{4,8\}$ bits per element, group size $G=64$, and a two-byte scale per group, the aligned low-bit payload is
\begin{equation}
 B_b(c)=cE\left(\frac{b}{8}+\frac{2}{G}\right)+2Sd.
 \label{eq:quantbytes}
\end{equation}
Here $G$ divides the quantized last dimension, as in the evaluated models. The BF16 boundary is independent of prefix precision. At 8K tokens, Qwen2.5-7B has a 56 MiB boundary and a 16 MiB BF16 KV layer. A 15-layer Q8 prefix therefore sends $15\cdot8.25+56=179.75$ MiB including scales.

\subsection{Overlap and cut selection}

Write $Q_b(c)$ for quantization time, $H$ for boundary transfer, $X_b(c)$ for prefix transfer and restoration, $R(c)$ for suffix execution, and $J$ for final synchronization. The sequential and ideal concurrent schedules have costs
\begin{align}
 T_{seq}(b,c)&=Q_b(c)+H+X_b(c)+R(c)+J, \\
 T_{ov}(b,c)&=Q_b(c)+H+\max\{X_b(c),R(c)\}+J.
 \label{eq:overlap}
\end{align}
The saved time is consequently
\begin{equation}
 T_{seq}(b,c)-T_{ov}(b,c)=\min\{X_b(c),R(c)\}.
\end{equation}
The shorter branch can be hidden entirely in this model. The boundary and quantization remain on the critical path. Executed timings include contention, restoration, and synchronization costs.

\Needspace{6\baselineskip}
\noindent\textbf{Proposition 2 (balanced cut in the linear overlap model).}
Fix the precision. Assume constant per-layer times $\alpha\geq0$ for quantization, $\beta>0$ for prefix transfer/restoration, and $\gamma>\alpha$ for suffix execution. Let $H$ and $J$ be independent of $c$. The minimizing integer interior cut is one of the two integers adjacent to
\begin{equation}
 c^*=\frac{L\gamma}{\beta+\gamma},
 \label{eq:balance}
\end{equation}
after clipping them to $[1,L-1]$.

\noindent\emph{Proof.} The branches intersect when $c\beta=(L-c)\gamma$, yielding~\eqref{eq:balance}. Substituting into~\eqref{eq:overlap} gives
\begin{equation}
 T(c)-H-J=
 \begin{cases}
 L\gamma+c(\alpha-\gamma), & c\leq c^*,\\
 c(\alpha+\beta), & c\geq c^*.
 \end{cases}
 \label{eq:piecewise}
\end{equation}
The first branch has negative slope and the second has positive slope. Hence the continuous minimum occurs at $c^*$. Every integer farther to the left has higher cost than $\lfloor c^*\rfloor$, and every integer farther to the right has higher cost than $\lceil c^*\rceil$. Restricting the interval clips these candidates to the nearest feasible endpoint. \hfill$\square$

If $\gamma\leq\alpha$, the left branch is nondecreasing and the smallest interior cut minimizes the model. Transfer-only and token-only full recomputation are evaluated separately because their boundary payloads differ from~\eqref{eq:quantbytes}. Increasing link bandwidth lowers the transfer component of $\beta$ and shifts $c^*$ toward more transmitted layers. These relations guide the cut sweep; the final policy uses measured latency.

\subsection{Quality constraint}

An action is $a=(b,c)$. We retain it when all evaluated next-token top-1 predictions agree with BF16, all evaluated needle-in-a-haystack (NIAH) prompts pass, and WikiText-2~\cite{wikitext} continuation perplexity rises by at most $\epsilon=0.01$ relative to BF16. With $\mathcal A_\epsilon$ denoting these eligible actions, selection is
\begin{equation}
 \widehat a=\underset{a\in\mathcal A_\epsilon}{\arg\min}\;
 \operatorname{median}_{i=1,\ldots,n} T_i(a).
 \label{eq:selection}
\end{equation}
The gate is evaluated on the fixed quality sets. BF16 and Q8 form the main comparison; Q4 expands the measured action space where it meets the same criterion.

The boundary construction localizes approximation to the transmitted prefix. To see this, let layer $\ell$ map its input $h_\ell$ to $(h_{\ell+1},K_\ell,V_\ell)$ with fixed weights, positions, and attention mask. Starting from the same $h_c$, the receiver obtains the same layer-$c$ outputs. Induction over $\ell=c,\ldots,L-1$ establishes equality of the recomputed suffix in exact arithmetic. The numerical implementation uses BF16 and is checked against the baseline; low-bit prefix restoration is the explicit source of approximation.

\section{Experimental Setup}

We evaluate the 3B and 7B Instruct variants of Qwen2.5~\cite{qwen25}, plus Qwen3-8B~\cite{qwen3}, with 512, 2048, and 8192 prompt tokens. The principal transfer experiment uses two separately allocated cross-node H100 GPUs on the measured cluster path. Each policy has three warmup iterations and 30 timed repetitions. The interval begins after prefill state is ready, before quantization, and ends after transfer, restoration, suffix execution, and the output projection for the first token. The transfer benchmark uses synthetic inputs and model-matched tensor dimensions: transferred tensors have the model's dimensions, and the suffix executes the model's layer dimensions, but the inputs and repeated layer weights in this pipeline are synthetic. The three policies compared on this path are full BF16 transfer, Q8 transfer-only, and the joint Q8 action; a separate BF16 suffix sweep supplies the recompute-only baseline. Reported latency is the median unless noted otherwise.

We also run a real-weight standalone P/D experiment from the beginning of prompt processing to the first greedy token returned by the decode GPU. This BF16 experiment times prefill, transfer, and token return, without Q8. The returned token uses suffix logits; the received prefix is not used for subsequent decoding. The quality runs use real weights and the same prefix quantizer: next-token distribution comparisons, fixed NIAH prompts, WikiText-2 continuation perplexity, and a fixed 64-question GSM8K~\cite{gsm8k} subset for Qwen2.5-3B and 7B. H200 and A100 are used for several quality checks, while H100 and L40S paths supply the transfer and crossover measurements. We compute 95\% confidence intervals by independent bootstrap resampling for separately executed handoff baselines and paired resampling for real-weight TTFT. Each timing point contains 30 repetitions.

\begin{figure*}[t]
  \centering
  \includegraphics[width=0.90\textwidth]{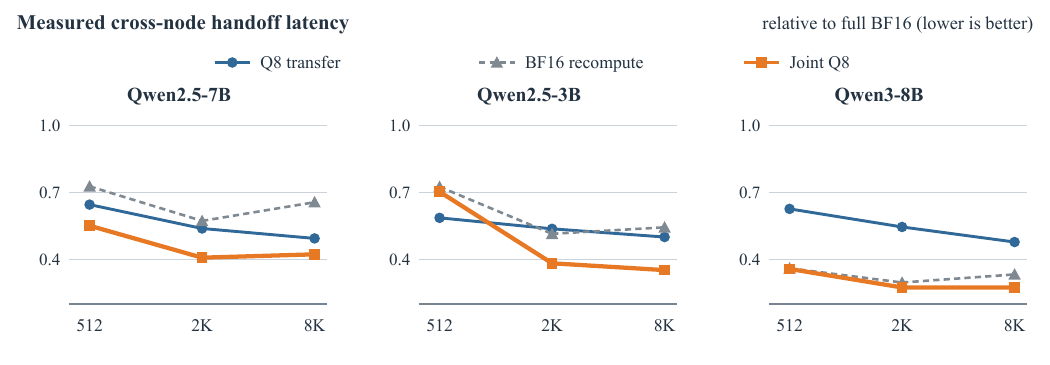}
  \caption{Executed post-prefill handoff latency normalized to BF16 full transfer on the cross-node H100 path (three warmup and 30 timed repetitions per action). The plot uses medians from the unified action sweep; the 8K values also appear in Table~\ref{tab:frontier}. Lower is better. The benchmark uses synthetic state and layer weights; Fig.~\ref{fig:ttft} reports the separate real-weight BF16 experiment.}
  \label{fig:latency}
\end{figure*}

\section{Results}

\subsection{Transfer latency under the quality constraint}

Figure~\ref{fig:latency} compares the configurations across prompt lengths, and Table~\ref{tab:frontier} gives their measured 8K latencies. Joint Q8 is faster than full BF16 transfer by 2.36$\times$, 2.84$\times$, and 3.65$\times$ for the 7B, 3B, and Qwen3-8B configurations. Relative to the faster of Q8 transfer-only and BF16 recompute-only, the joint action is 1.17$\times$, 1.42$\times$, and 1.22$\times$ faster. The 95\% bootstrap intervals for speedup over full BF16 are [2.33, 2.38], [2.56, 2.88], and [3.57, 3.71]. Table~\ref{tab:quality} reports output quality. The corresponding WikiText-2 perplexity changes are +0.077\%, $-0.075$\%, and +0.160\%; the evaluated NIAH sets pass 4/4 in each case. On the fixed 64-question GSM8K subset, joint Q8 matches BF16 at 46/64 for Qwen2.5-7B and 24/64 for Qwen2.5-3B.

\begin{table}[t]
\centering
\caption{Executed 8K handoff latency (median, ms) on the cross-node H100 path. ``Recomp.'' is the best measured BF16 suffix cut for that model and length. Joint Q8 uses suffix lengths 13, 17, and 21, respectively.}
\label{tab:frontier}
\setlength{\tabcolsep}{2.5pt}
\begin{tabular}{lrrrr}
\toprule
Model & BF16 & Q8 & Recomp. & Joint \\
\midrule
Qwen2.5-7B & 227.09 & 112.30 & 149.14 & \textbf{96.11} \\
Qwen2.5-3B & 144.14 & 72.18 & 78.45 & \textbf{50.82} \\
Qwen3-8B & 575.28 & 275.20 & 191.68 & \textbf{157.76} \\
\bottomrule
\end{tabular}
\end{table}

\begin{table}[t]
\centering
\caption{Real-weight quality at 8K for Q8 transfer with suffix recomputation. PPL change is relative to BF16. GSM8K entries show BF16 / ReKV correct answers on the same 64 questions; ``--'' denotes an unevaluated task.}
\label{tab:quality}
\setlength{\tabcolsep}{3pt}
\begin{tabular}{lrrr}
\toprule
Model & PPL change & NIAH & GSM8K \\
\midrule
Qwen2.5-7B & $+0.077\%$ & 4/4 & 46 / 46 \\
Qwen2.5-3B & $-0.075\%$ & 4/4 & 24 / 24 \\
Qwen3-8B & $+0.160\%$ & 4/4 & -- \\
\bottomrule
\end{tabular}
\end{table}

For Qwen2.5-3B at 512 tokens, Q8 transfer-only takes 4.77 ms, versus 5.72 ms for joint Q8. At 2048 tokens the ordering reverses: the best BF16 recompute-only action takes 18.85 ms and joint Q8 takes 14.02 ms. A fixed suffix policy would give up either short-context latency or the longer-context joint gain. Qwen3-8B has a smaller gap at 512 tokens: 11.38 ms for joint Q8 versus 11.47 ms for its best BF16 recompute-only action.

Qwen2.5-7B joint Q4 raises continuation perplexity by more than 100-fold at 8K and fails the quality criterion. Qwen3-8B joint Q4 passes at 8K but exceeds the 1\% perplexity threshold at 2048 tokens. The cross-model comparison therefore uses Q8.

At 8K, next-token KL divergence for Qwen2.5-7B decreases from $7.32\times10^{-3}$ with Q8 transfer alone to $1.52\times10^{-3}$ with suffix recomputation. This comparison measures the change in the output distribution in addition to continuation perplexity.

\subsection{Communication volume and executed overlap}

\begin{table}[t]
\centering
\caption{Qwen2.5-7B, 8K: communication and scheduling measurements. Payloads include scales and boundary state. The overlap comparison uses BF16 and a 13-layer suffix; its times exclude prefill.}
\label{tab:ablation}
\setlength{\tabcolsep}{4pt}
\begin{tabular}{lr}
\toprule
Payload configuration & MiB \\
\midrule
Full BF16 KV & 448.00 \\
Full Q8 KV & 231.00 \\
Q8 prefix + BF16 boundary & 179.75 \\
\midrule
BF16 suffix schedule & Median (ms) \\
\midrule
Sequential transfer and recomputation & 216.31 \\
Overlapping transfer and recomputation & 149.14 \\
\bottomrule
\end{tabular}
\end{table}

Table~\ref{tab:ablation} separates communication volume from the effect of overlap for Qwen2.5-7B at 8K. Full BF16 sends 448.00 MiB, full Q8 sends 231.00 MiB including scales, and the Q8 prefix with a 13-layer suffix sends 179.75 MiB including the BF16 boundary state. All 30 measured receives for each payload complete without a recorded error. In a separate executed BF16 suffix run at the same cut, sending the boundary first and overlapping suffix execution with prefix transfer reduces the median transfer-and-recomputation time from 216.31 to 149.14 ms, a 1.45$\times$ schedule gain. This BF16 run isolates overlap; the joint Q8 point in Table~\ref{tab:frontier} includes the low-bit prefix as well.

\subsection{First-token latency and concurrency}

\begin{figure}[t]
\centering
\includegraphics[width=\columnwidth]{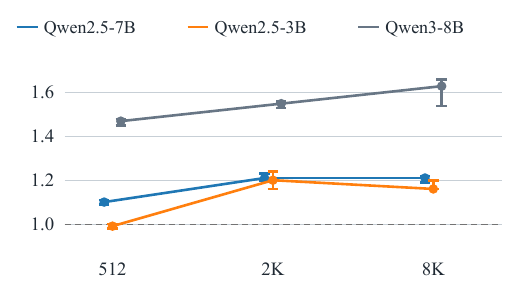}
\caption{Real-weight BF16 suffix-recomputation speedup over full transfer, measured from prefill to token return. Bars show paired-bootstrap 95\% confidence intervals from 30 repetitions. The horizontal dashed line denotes equal TTFT.}
\label{fig:ttft}
\end{figure}

Figure~\ref{fig:ttft} reports the separate real-weight BF16 experiment. At 8K, overlapping BF16 transfer and suffix execution reduces prefill-to-token-return time from 449.74 to 371.86 ms for Qwen2.5-7B, from 251.10 to 215.61 ms for Qwen2.5-3B, and from 853.43 to 523.89 ms for Qwen3-8B. These are 1.21$\times$ [1.19, 1.22], 1.16$\times$ [1.16, 1.20], and 1.63$\times$ [1.54, 1.66] speedups with 95\% bootstrap intervals. The timed interval includes prompt execution. All 30 returned tokens match the BF16 baseline; this checks suffix logits, not decoding with a reconstructed Q8 cache.

For simultaneously ready Qwen2.5-7B requests on two H100s, joint Q8 improves throughput over Q8 transfer-only for the measured batch sizes 2 and 4. At 8K and batch size 4, batch completion falls from 420.53 to 379.15 ms, corresponding to 9.51 versus 10.55 requests/s. The test uses a standalone contiguous cache and simultaneously ready requests.

\subsection{Link-dependent crossover}

The measured L40S path delivers roughly 20 GB/s for the 8K full-KV transfer. Under that path and the measured L40S layer cost, Qwen2.5-7B selects full BF16 transfer; for Llama-3.1-8B the best modeled nonzero cut improves latency by only 1.004$\times$. On the slower cross-node H100 TCP path, the 8K Qwen2.5-7B full BF16 payload of 448 MiB takes 202.13 ms median, making local suffix work more competitive. The L40S comparison uses a path profile and modeled cut selection; the measured comparison under the quality constraint uses the H100 path.

\section{Related Work and Scope}

PagedAttention and Sarathi-Serve address KV placement and prefill/decode scheduling within serving engines~\cite{vllm,sarathi}. Splitwise and DistServe separate the phases across workers~\cite{splitwise,distserve}; Mooncake and D\'ej\`aVu organize KV state and streaming across devices~\cite{mooncake,dejavu}. KIVI and KVQuant study low-bit KV representation~\cite{kivi,kvquant}, while CacheGen compresses and streams cached context over a bandwidth-varying path~\cite{cachegen}. SpectrumKV allocates transfer precision across tokens in P/D serving~\cite{spectrumkv}. CacheBlend and KVPR overlap partial cache recomputation with cache retrieval~\cite{cacheblend,kvpr}, while CacheFlow schedules restoration across token, layer, and GPU dimensions~\cite{cacheflow}. ReKV's measured comparison concerns a fresh GPU--GPU P/D handoff and the quality-constrained choice between low-bit prefix transfer and a lossless contiguous suffix.

Other systems organize where and when KV state moves. MemServe pools reusable context, Llumnix migrates active requests, P/D-Serve scales phase-separated deployments, and KVDirect optimizes inter-node transfer~\cite{memserve,llumnix,pdserve,kvdirect}. ServerlessLLM supports live migration; HCache restores state from intermediate activations; LayerKV manages allocation and offloading per layer~\cite{serverless,hcache,layerkv}. KVServe selects communication-compression profiles online, HACK computes on compressed KV, and SmartGen transfers selected entries~\cite{kvserve,hack,smartgen}. SGLang reuses prefix state across generation calls~\cite{sglang}.

\section{Conclusion}

On the measured slow cross-node H100 path, a quantized prefix and recomputed suffix yield transfer latencies meeting the quality criteria that neither Q8 transfer alone nor BF16 recomputation alone reaches at 8K tokens for three models. The preferred action changes at short prompts and on a faster link. A separate real-weight BF16 experiment improves prefill-to-token-return time by 1.16--1.63$\times$ at 8K. End-to-end Q8 decoding and paged-cache serving integration remain to be evaluated.


\clearpage
\begin{thebibliography}{30}
\bibitem{sarathi} A. Agrawal \emph{et al.}, ``Taming throughput-latency tradeoff in LLM inference with Sarathi-Serve,'' in \emph{USENIX OSDI}, 2024. \href{https://www.usenix.org/conference/osdi24/presentation/agrawal}{[link]}
\bibitem{splitwise} P. Patel \emph{et al.}, ``Splitwise: Efficient generative LLM inference using phase splitting,'' arXiv:2311.18677, 2023. \href{https://arxiv.org/abs/2311.18677}{[link]}
\bibitem{distserve} Y. Zhong \emph{et al.}, ``DistServe: Disaggregating prefill and decoding for goodput-optimized large language model serving,'' in \emph{USENIX OSDI}, 2024. \href{https://www.usenix.org/conference/osdi24/presentation/zhong-yinmin}{[link]}
\bibitem{mooncake} R. Qin \emph{et al.}, ``Mooncake: Trading More Storage for Less Computation---A KVCache-centric Architecture for Serving LLM Chatbot,'' in \emph{USENIX FAST}, 2025. \href{https://www.usenix.org/conference/fast25/presentation/qin}{[link]}
\bibitem{dejavu} F. Strati \emph{et al.}, ``D\'ej\`aVu: KV-cache streaming for fast, fault-tolerant generative LLM serving,'' in \emph{ICML}, 2024. \href{https://arxiv.org/abs/2403.01876}{[link]}
\bibitem{kivi} Z. Liu \emph{et al.}, ``KIVI: A tuning-free asymmetric 2bit quantization for KV cache,'' in \emph{ICML}, 2024. \href{https://arxiv.org/abs/2402.02750}{[link]}
\bibitem{kvquant} C. Hooper \emph{et al.}, ``KVQuant: Towards 10 million context length LLM inference with KV cache quantization,'' in \emph{NeurIPS}, 2024. \href{https://proceedings.nips.cc/paper_files/paper/2024/hash/028fcbcf85435d39a40c4d61b42c99a4-Abstract-Conference.html}{[link]}
\bibitem{cachegen} Y. Liu \emph{et al.}, ``CacheGen: KV cache compression and streaming for fast large language model serving,'' in \emph{ACM SIGCOMM}, 2024. \href{https://doi.org/10.1145/3651890.3672274}{[link]}
\bibitem{spectrumkv} P. Yang, ``SpectrumKV: Per-token mixed-precision KV cache transfer for prefill-decode disaggregated LLM serving,'' arXiv:2606.08635, 2026. \href{https://arxiv.org/abs/2606.08635}{[link]}
\bibitem{kvpr} C. Jiang \emph{et al.}, ``KVPR: Efficient LLM inference with I/O-aware KV cache partial recomputation,'' in \emph{Findings of ACL}, 2025. \href{https://aclanthology.org/2025.findings-acl.997/}{[link]}
\bibitem{cacheflow} S. Nian \emph{et al.}, ``CacheFlow: Efficient LLM serving with 3D-parallel KV cache restoration,'' arXiv:2604.25080, 2026. \href{https://arxiv.org/abs/2604.25080}{[link]}
\bibitem{gqa} J. Ainslie \emph{et al.}, ``GQA: Training generalized multi-query transformer models from multi-head checkpoints,'' in \emph{EMNLP}, 2023. \href{https://aclanthology.org/2023.emnlp-main.298/}{[link]}
\bibitem{qwen25} A. Yang \emph{et al.}, ``Qwen2.5 technical report,'' arXiv:2412.15115, 2024. \href{https://arxiv.org/abs/2412.15115}{[link]}
\bibitem{llama3} A. Grattafiori \emph{et al.}, ``The Llama 3 herd of models,'' arXiv:2407.21783, 2024. \href{https://arxiv.org/abs/2407.21783}{[link]}
\bibitem{wikitext} S. Merity \emph{et al.}, ``Pointer sentinel mixture models,'' arXiv:1609.07843, 2016. \href{https://arxiv.org/abs/1609.07843}{[link]}
\bibitem{qwen3} A. Yang \emph{et al.}, ``Qwen3 technical report,'' arXiv:2505.09388, 2025. \href{https://arxiv.org/abs/2505.09388}{[link]}
\bibitem{gsm8k} K. Cobbe \emph{et al.}, ``Training verifiers to solve math word problems,'' arXiv:2110.14168, 2021. \href{https://arxiv.org/abs/2110.14168}{[link]}
\bibitem{vllm} W. Kwon \emph{et al.}, ``Efficient memory management for large language model serving with PagedAttention,'' in \emph{ACM SOSP}, 2023. \href{https://doi.org/10.1145/3600006.3613165}{[link]}
\bibitem{cacheblend} J. Yao \emph{et al.}, ``CacheBlend: Fast large language model serving for RAG with cached knowledge fusion,'' in \emph{ACM EuroSys}, 2025. \href{https://arxiv.org/abs/2405.16444}{[link]}
\bibitem{memserve} C. Hu \emph{et al.}, ``MemServe: Context caching for disaggregated LLM serving with elastic memory pool,'' arXiv:2406.17565, 2024. \href{https://arxiv.org/abs/2406.17565}{[link]}
\bibitem{llumnix} B. Sun \emph{et al.}, ``Llumnix: Dynamic scheduling for large language model serving,'' in \emph{USENIX OSDI}, 2024. \href{https://www.usenix.org/conference/osdi24/presentation/sun-biao}{[link]}
\bibitem{pdserve} Y. Jin \emph{et al.}, ``P/D-Serve: Serving disaggregated large language model at scale,'' arXiv:2408.08147, 2024. \href{https://arxiv.org/abs/2408.08147}{[link]}
\bibitem{kvdirect} S. Chen \emph{et al.}, ``KVDirect: Distributed disaggregated LLM inference,'' arXiv:2501.14743, 2025. \href{https://arxiv.org/abs/2501.14743}{[link]}
\bibitem{serverless} Y. Fu \emph{et al.}, ``ServerlessLLM: Low-latency serverless inference for large language models,'' in \emph{USENIX OSDI}, 2024. \href{https://www.usenix.org/conference/osdi24/presentation/fu}{[link]}
\bibitem{hcache} S. Gao \emph{et al.}, ``Fast state restoration in LLM serving with HCache,'' in \emph{ACM EuroSys}, 2025. \href{https://arxiv.org/abs/2410.05004}{[link]}
\bibitem{layerkv} Y. Xiong \emph{et al.}, ``LayerKV: Optimizing large language model serving with layer-wise KV cache management,'' arXiv:2410.00428, 2024. \href{https://arxiv.org/abs/2410.00428}{[link]}
\bibitem{kvserve} Z. Liu \emph{et al.}, ``KVServe: Service-aware KV cache compression for communication-efficient disaggregated LLM serving,'' arXiv:2605.13734, 2026. \href{https://arxiv.org/abs/2605.13734}{[link]}
\bibitem{hack} Z. Zhang \emph{et al.}, ``HACK: Homomorphic acceleration via compression of the key-value cache for disaggregated LLM inference,'' arXiv:2502.03589, 2025. \href{https://arxiv.org/abs/2502.03589}{[link]}
\bibitem{smartgen} X. Luo \emph{et al.}, ``SmartGen: Seamless disaggregated LLM inference with selective KV cache transfer,'' arXiv:2607.28150, 2026. \href{https://arxiv.org/abs/2607.28150}{[link]}
\bibitem{sglang} L. Zheng \emph{et al.}, ``SGLang: Efficient execution of structured language model programs,'' in \emph{NeurIPS}, 2024. \href{https://arxiv.org/abs/2312.07104}{[link]}
\end{thebibliography}
\end{document}